\documentstyle[preprint,prb,epsfig,aps]{revtex}
\begin{document}
\draft
\title{Determination of superconducting anisotropy from
magnetization data on random powders as applied to LuNi$_2$B$_2$C,
YNi$_2$B$_2$C and MgB$_2$.}
\author{S. L. Bud'ko, V. G. Kogan, and P. C. Canfield}
\address{Ames Laboratory and Department of
Physics and Astronomy, Iowa State University, Ames, Iowa 50011}

\date{\today}
\maketitle
\begin{abstract}

The recently discovered intermetallic superconductor
MgB$_2$ appears to have a highly anisotopic upper critical field with
$ H_{c2}^{max}/H_{c2}^{min} \equiv \gamma  > 5$.  In order to determine
the temperature dependence of  both $H_{c2}^{max}$ 
and $H_{c2}^{min}$ 
we propose a method of extracting the superconducting anisotropy
from the magnetization $M(H,T)$ of randomly oriented powder samples.
The method is based on  two features in
$(\partial M/\partial T)_H$:
the onset of diamagnetism at $T_c^{max}$, that is
commonly  associated with $H_{c2}$, and a  kink in
$\partial M/\partial T$ at a lower temperature $T_c^{min}$.
Results for LuNi$_2$B$_2$C and
YNi$_2$B$_2$C powders are in agreement with anisotropic $H_{c2}$
  obtained from magneto-transport measurements on single crystals.  Using
this method on four different types of  MgB$_2$ powder samples we are able to
determine $H_{c2}^{max}(T)$ and
$H_{c2}^{min}(T)$   with $\gamma \approx 6$.
\\
\end{abstract}


Anisotropic type-II superconductors have been studied for a number of
years\cite{NbSe2} with a revisited interest  after the discovery   of
strongly anisotropic high-$T_c$ cuprates. The possibility has
recently been raised that MgB$_2$ ($T_c \approx 40$ K)
may also be highly anisotropic.\cite{ferenc}
The quantitative characterisation of the anisotropy  is usually
one of the first questions 
to be addressed after the material is
synthesized.
One of the most accurate methods of extracting the anisotropy parameter
$\gamma = H_{c2}^{max}/H_{c2}^{min}$ is by measuring the torque
upon the single crystal sample in
intermediate fields
$H_{c1}\ll H \ll H_{c2}$ inclined relative to the crystal axes
(we are interested here in uniaxial materials  with
large Ginzburg-Landau (GL) parameter $\kappa=\lambda /\xi$;
$\lambda$ and $\xi$  are the penetration depth and the coherence
length, $H_{c1}$ and  $H_{c2}$ are the lower and the upper critical
fields).\cite{YBCO} The  method is based on the existence of the
transverse magnetization in anisotropic superconductors which causes the
torque, the angular dependence of which depends only on
$\gamma$.\cite{K88}

Often,  materials are first synthesized in a polycrystalline form.
  Anisotropy estimates of such samples can be inferred by a number
of measurements.
In one of the common schemes, \cite{farrell} a powder of the
material is mixed with a low magnetic background epoxy and aligned in a high
   magnetic field. The alignment is made permanent by curing the epoxy. The
necessary requirements include a powder that consists of single crystalline
grains  with a considerable  normal state magnetic anisotropy. The samples
obtained in this way are suitable for anisotropic magnetization measurements.
Usually, uncertainties in a degree of alignment lead to underestimates of
$\gamma$.

We describe here a simple  method  to assess the anisotropy
    and to delineate  $H_{c2}^{max}(T)$ and $H_{c2}^{min}(T)$
by analysis of the magnetization  $M(T)$ taken for several values
of the applied  field $H$.  We   establish  the validity of this method by
comparison of anisotropic
$H_{c2}(T)$ measured  directly (by magneto-transport) on single
crystals of quaternary borocarbide  superconductors\cite{PhysToday}
LuNi$_2$B$_2$C and YNi$_2$B$_2$C  with the  values extracted from the
$M(H,T)$ data on randomly oriented powders. We then
use the method to determine the curves of $H_{c2}^{max}(T)$ and
$H_{c2}^{min}(T)$   for MgB$_2$ and to obtain
$\gamma \approx 6$, close to the values given in Ref. \onlinecite{ferenc}.

Single crystals of LuNi$_2$B$_2$C and YNi$_2$B$_2$C   were
grown using the Ni$_2$B
flux growth technique.\cite{PhysToday,MXu} Four different samples of
MgB$_2$ were used in this work: (i) polycrystalline sintered
Mg$^{11}$B$_2$,   (ii) MgB$_2$ wire segments,
and commercial 99.5\% purity  MgB$_2$ powders from two sources:  (iii)
Accumet Materials Co. and  (iv) Alfa Aesar. Samples (i) and (ii) were
synthesized by reacting elemental boron with Mg
at $950^{\circ}$ C.\cite{iso,DKF,AH}

The anisotropic $H_{c2}(T)$  of LuNi$_2$B$_2$C and YNi$_2$B$_2$C
single crystals were measured directly  by monitoring
resistance $R(H,T)$ near the
superconducting transition.  For the magnetization  measurements,
single crystals of borocarbides and  bulk samples of  MgB$_2$ were
ground to powder,  mixed with Epocast 121 epoxy  (the epoxy has weak
featureless diamagnetism in the $H-T$ range of interest),
stirred to assure random orientation of particles and cured at $120^{\circ}$ C
for 1 hour.
DC magnetization measurements were performed in Quantum Design MPMS-5 or
MPMS-7 SQUID
magnetometers. Resistance of borocarbides were measured
using LR-700 AC resistance bridge ($f$ = 16 Hz, $I$ = 1-3 mA) and the $H-T$
environment of the MPMS instruments. Contacts to the samples were made
with Epo-tek H20E silver epoxy. Currents were
flowing in the $ab$ plane, whereas the  field was applied
perpendicular to  the current, along $c$ axis or in the $ab$ plane.

\begin{figure}
\epsfxsize=0.8\hsize
\centerline{
\vbox{
\epsffile{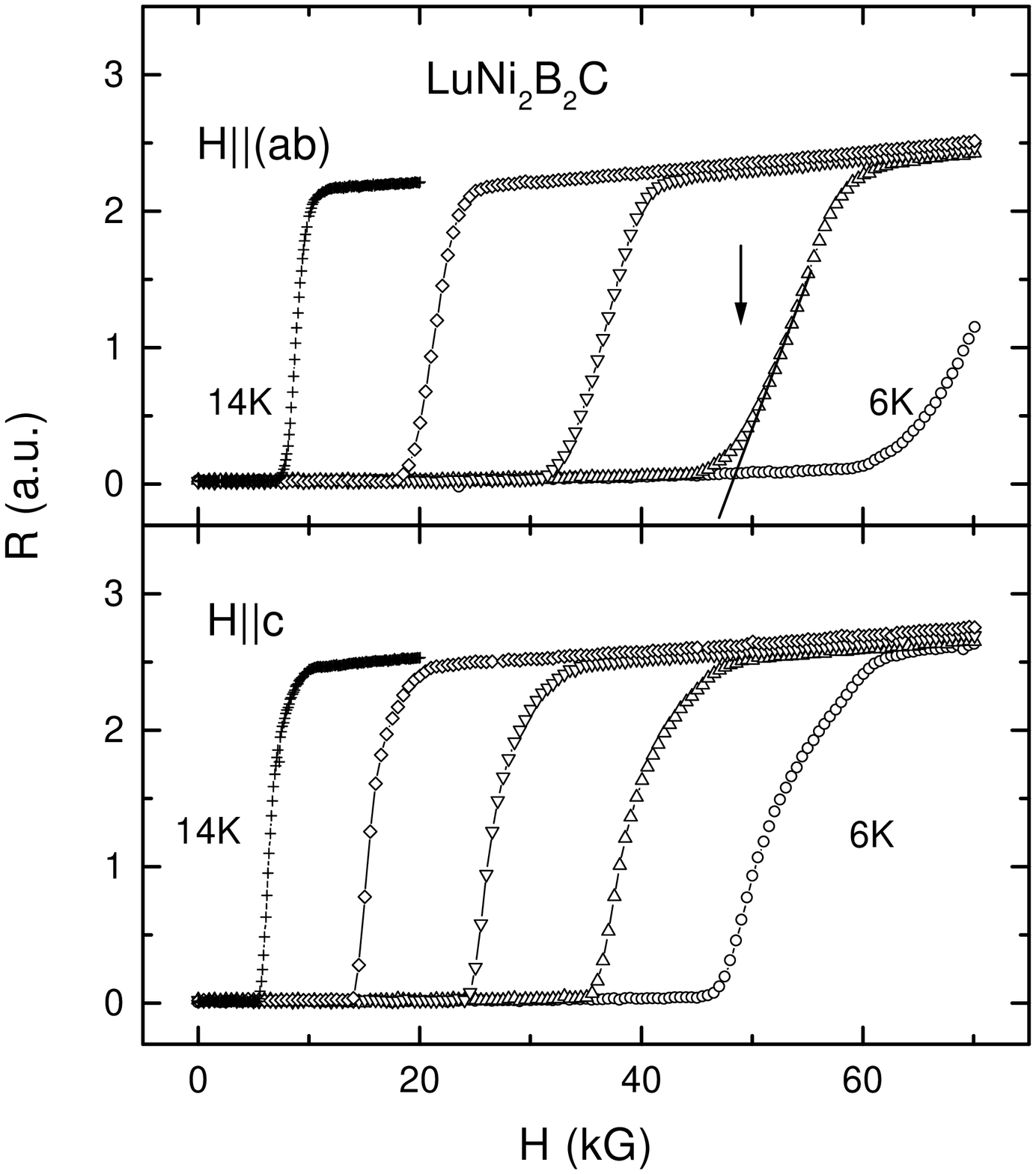}
}}
\vskip \baselineskip
\caption{Representative $R(H)$ curves for LuNi$_2$B$_2$C. Data shown are
from 6 K to 14 K taken every 2 K. The criterion for $H_{c2}(T)$
is shown by the vertical arrow in the upper panel.}
\label{FLuR}
\end{figure}

The data on anisotropic $H_{c2}$ for non-magnetic  borocarbides were
reported in several publications.\cite{MX1,Tq,TAM,Argonne}
Since the $H_{c2}$  anisotropy appears to be
somewhat sample-dependent, we  performed direct
magneto-transport measurements on   samples from the
same batches that were used for random powder magnetization
measurements.  Figure \ref{FLuR} shows representative
resistance, $R(H)$, data  for  LuNi$_2$B$_2$C
measured at constant temperatures
for two field  orientations. The resistively measured
superconducting  transition broadens slightly in field; this results in
criterion-dependent $H_{c2}(T)$ curves. Still, the anisotropy of $H_{c2}(T)$
turns out  practically criterion-independent. In the following we
use the criterion of the maximum  slope intersection with the $R = 0$
line, as shown in Fig.\ref{FLuR}.

\begin{figure}
\epsfxsize=0.9\hsize
\vbox{
\centerline{
\epsffile{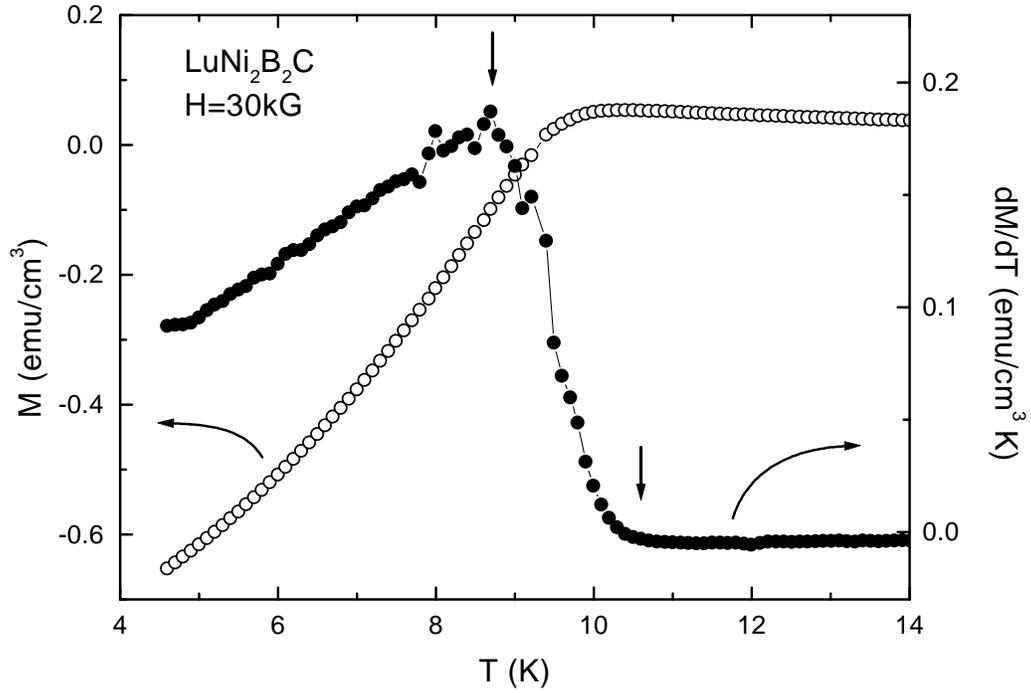}
}}
\caption{Magnetization  $M(T)$  for $H = 30$ kG   and its temperature
derivative for a randomly oriented  powder of LuNi$_2$B$_2$C.
The vertical arrows show $T_c^{min}(H)$ and $T_c^{max}(H)$.}
\label{FLuMder}
\end{figure}

An example of magnetization data, $M(T)$,  in a constant
   field, $H = 30$ kG, for a powder sample of LuNi$_2$B$_2$C is
shown in Fig. \ref{FLuMder} along with the temperature derivative
$\partial M/\partial T$.
We interpret the behavior of $\partial M/\partial T$ as follows.
Upon cooling a powdered
sample in a fixed $H$, there is   a deviation
from a roughly $T$ independent normal state magnetization
(for non-magnetic materials as is the case for the
three compounds discussed here) to an increasingly diamagnetic
signal at $T = T_c^{max}$.
A second sharp 
feature will occur in $\partial M/\partial T$ when the sample 
temperature passes through $T_c^{min}$.  This can be most clearly 
understood by considering what happens to the sample upon warming.  
For $T < T_c^{min}$ all grains are superconducting 
whereas for $T > T_c^{min}$ part of them will become normal, 
depending upon their orientation with respect to the applied magnetic 
field.  Therefore, upon warming through $T_c^{min}$ there will be a 
kink in $\partial M/\partial T$ associated with the onset of normal 
state properties in an increasing number of appropriately oriented 
grains.  In Fig. \ref{FLuMder} this can be seen in the 
$\partial M/\partial T$ plot with $T_c^{max} =  10.6$ K and 
$T_c^{min} = 8.7$ K. The temperatures $T_c^{min}$ and
$T_c^{max}$ are  marked with vertical  arrows in Fig. \ref{FLuMder}.
It is worth noting that  when a single $H_{c2}(T)$  value is obtained 
from measurements
on polycrystalline samples it is actually the maximum upper critical field
$H_{c2}^{max}(T)$. \cite{DKF}  

\begin{figure}
\epsfxsize=0.9\hsize
\vbox{
\centerline{
\epsffile{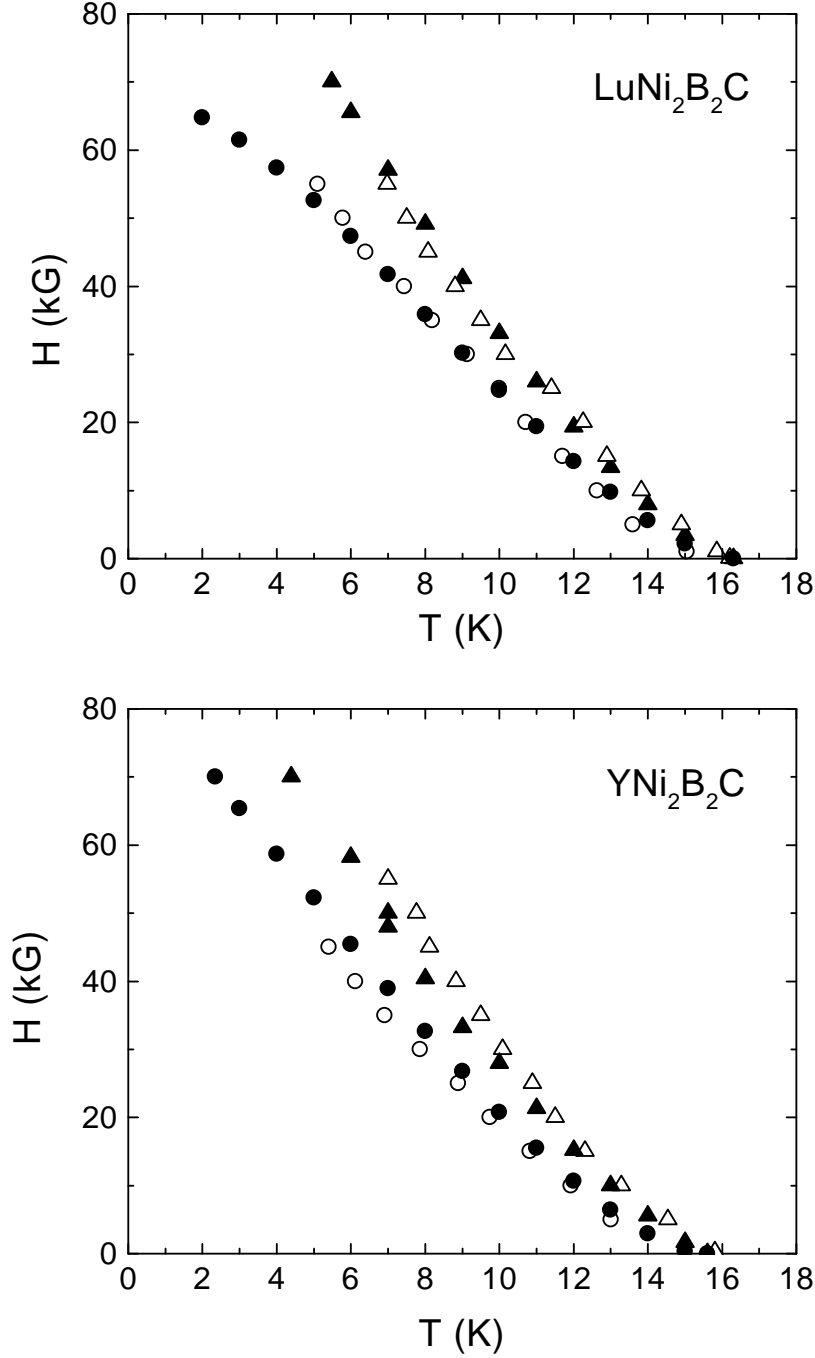}
}}
\caption{The upper panel: the minimum upper critical field $H_{c2}^{min}(T)$
(open circles) and the maximum one $H_{c2}^{max}(T)$ (open  triangles) for
LuNi$_2$B$_2$C powder obtained by analysis of $\partial M/\partial T$.   Filled
circles and  triangles are $H_{c2}^c(T)$ and
$H_{c2}^{ab}(T)$ respectively from magneto-resistance measurements on
LuNi$_2$B$_2$C crystals. The lower panel: the same for YNi$_2$B$_2$C}
\label{FLuYH}
\end{figure}

The $M(T)$ data   taken for a  number of fixed   fields were
analyzed in a similar  manner; the resulting curves $H_{c2}^{min}(T)$
and $H_{c2}^{max}(T)$  are  presented in the upper panel of 
Fig. \ref{FLuYH} (open symbols).
These are
plotted together with  the directly measured anisotropic
$H_{c2}(T)$  from
magneto-transport measurements  on LuNi$_2$B$_2$C single crystals from
the same batch.  It is seen that both sets of data are in
agreement. We obtain  $\gamma \approx 1.2 -
1.3$.   Albeit slightly higher, this value of $\gamma$ is consistent
with values reported in   literature: e.g. $\gamma \approx 1.2$.
\cite{Argonne}
Similar measurements and analysis were performed on YNi$_2$B$_2$C. The results
are shown in lower panel of Fig. \ref{FLuYH}.
The slightly higher mismatch between the magneto-transport $H_{c2}$'s of single
crystals and those extracted from powder magnetization measurements
than for the case of LuNi$_2$B$_2$C might
be due to arbitrary in-plane field orientation in magneto-transport
measurements, choice of criteria and/or error bars in the measurements.

It is worth noting that the method of extracting the 
anisotropic $H_{c2}(T)$ and the anisotropy parameter
$\gamma$  from the powder data  presented here constitutes a robust
procedure independent of a particular model for describing the anisotropy.
Moreover, the method is just based  on the existence of a  kink in
$\partial M/\partial T$ located at $T = T_c^{min}$, a feature which should be
present for any angular distribution of the grains (as long as it is 
continuous,
but not necessarily random). The analysis of the $M(H,T)$ data   can be pushed
further to to relate  $\gamma$ to other material
characteristics ($\kappa$, $\lambda$, etc). \cite{KC,GKL} This, however, would
have involved more information (e.g., randomness, assumptions on linear
dependences of $M$ on $H_{c2}-H$ and $T_c-T$) and we will not pursue this here.

\begin{figure}
\epsfxsize=0.9\hsize
\vbox{
\centerline{
\epsffile{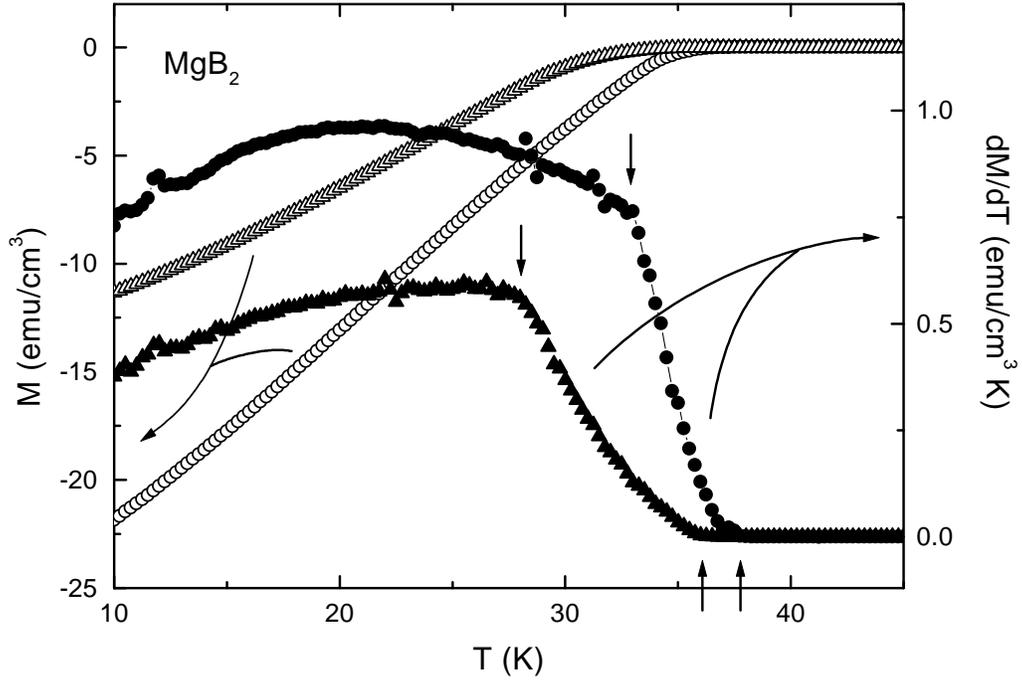}
}}
\caption{Magnetization $M(T)$ of the powder sample of MgB$_2$ at
$H = 5$ kG (open circles) and  $10$ kG (open triangles) and their temperature
derivatives (respective filled symbols).
The vertical arrows show $T_c^{min}(H)$ and $T_c^{max}(H)$ for each field.}
\label{FMgMder}
\end{figure}

With the validity of this method
established for borocarbides, we can use it for
MgB$_2$.\cite{jap}
An example of  magnetization curves for two   values of $H$
and their temperature derivatives are shown in Fig. \ref{FMgMder}. As in
borocarbides, a clear feature
at $T_c^{min}$ is
seen in $\partial M/\partial T$.   With increasing $H$, this feature moves down
in temperature  faster than $T_c^{max}$ and disappears for $H > 25$ kG.

\begin{figure}
\epsfxsize=0.9\hsize
\vbox{
\centerline{
\epsffile{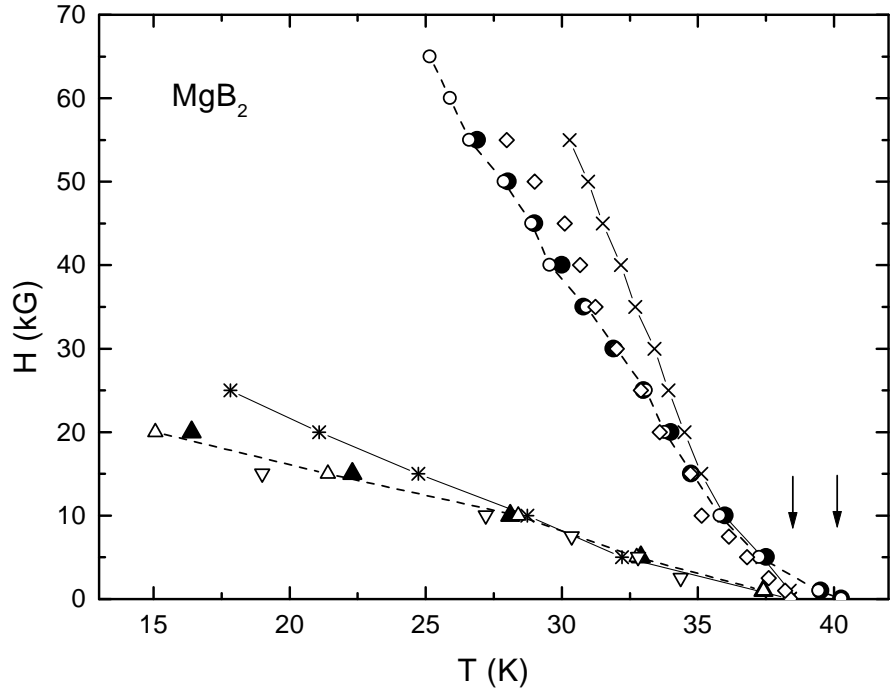}
}}
\caption{Anisotropic $H_{c2}(T)$ curves for four different samples of
MgB$_2$ obtained from the analysis of magnetization. Filled circles
and up triangles - sintered powder, open circles
and up triangles - wire segments, crosses and astericks - Accumet, open
diamonds and down triangles - Alfa Aesar. The right vertical arrow
shows $T_c$ for sintered powder and wires, the left one, for commercial
powders.}
\label{MgB2H}
\end{figure}

$H_{c2}^{min}(T)$ and $H_{c2}^{max}(T)$ curves were
deduced from magnetization data collected in fixed
   fields for four different samples of MgB$_2$. These data
are presented in Fig. \ref{MgB2H}.
Data for the sintered powder and the
powdered wire segments are very similar. The $H_{c2}^{max}(T)$
curves are consistent with the reported \cite{DKF,AH} polycrystalline
magneto-transport $H_{c2}(T)$.  The extracted anisotropy is $\gamma \approx
5-6$. Both commercial MgB$_2$ powders have $T_c$ values that are 1-1.5 K lower than 
sintered powder
or wire segments (see Fig. \ref{MgB2H}), presumably due to a higher levels
of impurities. The features in  $\partial M/\partial T$
are less pronounced
in commercial powders, however the
resulting  anisotropies are similar to that observed in clean MgB$_2$ (Fig.
\ref{MgB2H}).  For these samples $\gamma \approx 6-7$.

There have been
some initial
reports of the  $H_{c2}$ anisotropy for
MgB$_2$.  The anisotropy  $\gamma$  for separate particles
settled on a flat surface is 1.73.  \cite{lima}  For a hot-pressed bulk
sample,   $\gamma\approx 1.1$. \cite{dresden}  Measurements on $c$-axis
oriented thin films \cite{madison} gave   $\gamma =
1.8 - 2$, with the higher $\gamma$ for  films with higher resistivity
and lower $T_c$. Recently, there were  several announcements on growth and
anisotropic properties of sub-mm sized  single crystals of MgB$_2$.
\cite{Xtal1,Xtal2,Xtal3} In all three cases  the residual resistance ratio
was $5 - 7$ and the   transition  temperature was
$\approx 0.5 - 1$ K lower  than for polycrystalline materials.
\cite{iso,AH,jap} The $H_{c2}$  anisotropy was found to
be in the range $2.6 - 3$. A significantly higher $H_{c2}$
anisotropy ($\gamma \approx 6 - 9$)  was inferred
from conduction electron spin resonance  measurements and from fits to the reversible
part of $M(H)$ on high purity and high residual resistance ratio samples
\cite{ferenc}. The results for MgB$_2$ obtained in this work are consistent
with the latter and are at the higher end of the  wide-spread 
of values in the literature.

The method of extracting the superconducting anisotropy suggested in this
paper works well for random powders of LuNi$_2$B$_2$C and YNi$_2$B$_2$C, for
which the results were compared with direct measurements of $H_{c2}$.
Moreover, the
$\gamma$ values so obtained are in agreement with the
microscopic theory\cite{Gor'kov} according to which
\begin{equation}
\gamma^2= {\langle\Delta({\bf k}_F)v_{ab}^2\rangle\over \langle\Delta({\bf
k}_F)v_c^2\rangle}\,, \label{gorkov}
\end{equation}
where $v_i$ are  the
Fermi velocities and   $\langle...\rangle$ stand for Fermi surface
averages.
For the isotropic gap function $\Delta({\bf k}_F)=$ constant, the 
band structure
estimates give: \cite{harmon,dugdale}
\begin{equation}
\gamma = \sqrt{\langle v_{ab}^2\rangle / \langle
   v_c^2\rangle }\approx 1.2\,\,.
\end{equation}

This, however, is not the case for MgB$_2$. The ratio
$\langle v_{ab}^2\rangle / \langle v_c^2\rangle$ averaged over the {\it
whole} Fermi surface  for this material is  close to
unity,\cite{Kirill}  whereas our values of $\gamma^2$
are in the range  $25 - 50$. This is indicative of a strong
anisotropy of $\Delta({\bf k}_F)$. There are 
arguments \cite {Liu}  that the electron-phonon interaction is particularly
strong on the Fermi surface sheets shaped as slightly distorted 
cylinders along the $c$
crystal direction.  The ratio $\langle v_{ab}^2\rangle / \langle
v_c^2\rangle$ averaged only over these cylinders is estimated as $\approx
40$.\cite{Kir-And} If the gap $\Delta$ on the remaining Fermi surface sheets
is negligible, we expect the superconducting anisotropy as $\gamma \approx
\sqrt{40} \approx 6$, the value close to that extracted from our analysis of the
magnetization.

In conclusion, we suggest  a simple method of evaluation of the anisotropy
of the   upper critical field from the analysis of
the temperature dependent magnetization of randomly oriented
powders. In the case of
non-magnetic borocarbides the results are in good agreement with the direct
measurements on single crystals and with the band-structure calculations
provided the gap is isotropic. For MgB$_2$ the estimated anisotropy is
$\gamma \approx 5 - 7$, which can be reconciled with the band calculations
only if the superconducting gap on the cylindrically shaped sheets of the
Fermi surface is dominant.
\\

We thank K. D. Belashchenko and V. P. Antropov for discussions and 
for providing
the $\langle v_{ab}^2\rangle / \langle v_c^2\rangle$ value for cylindrical
parts of MgB$_2$ Fermi surface.
Ames Laboratory is operated for the
U. S. Department of Energy by Iowa State University under contract No.
W-7405-ENG-82. This work was
supported by the Director of Energy Research, Office of Basic Energy Sciences.

%
\pagebreak
\pagebreak
\pagebreak
\begin{figure}
\epsfxsize=0.8\hsize
\vbox{
\centerline{
\epsffile{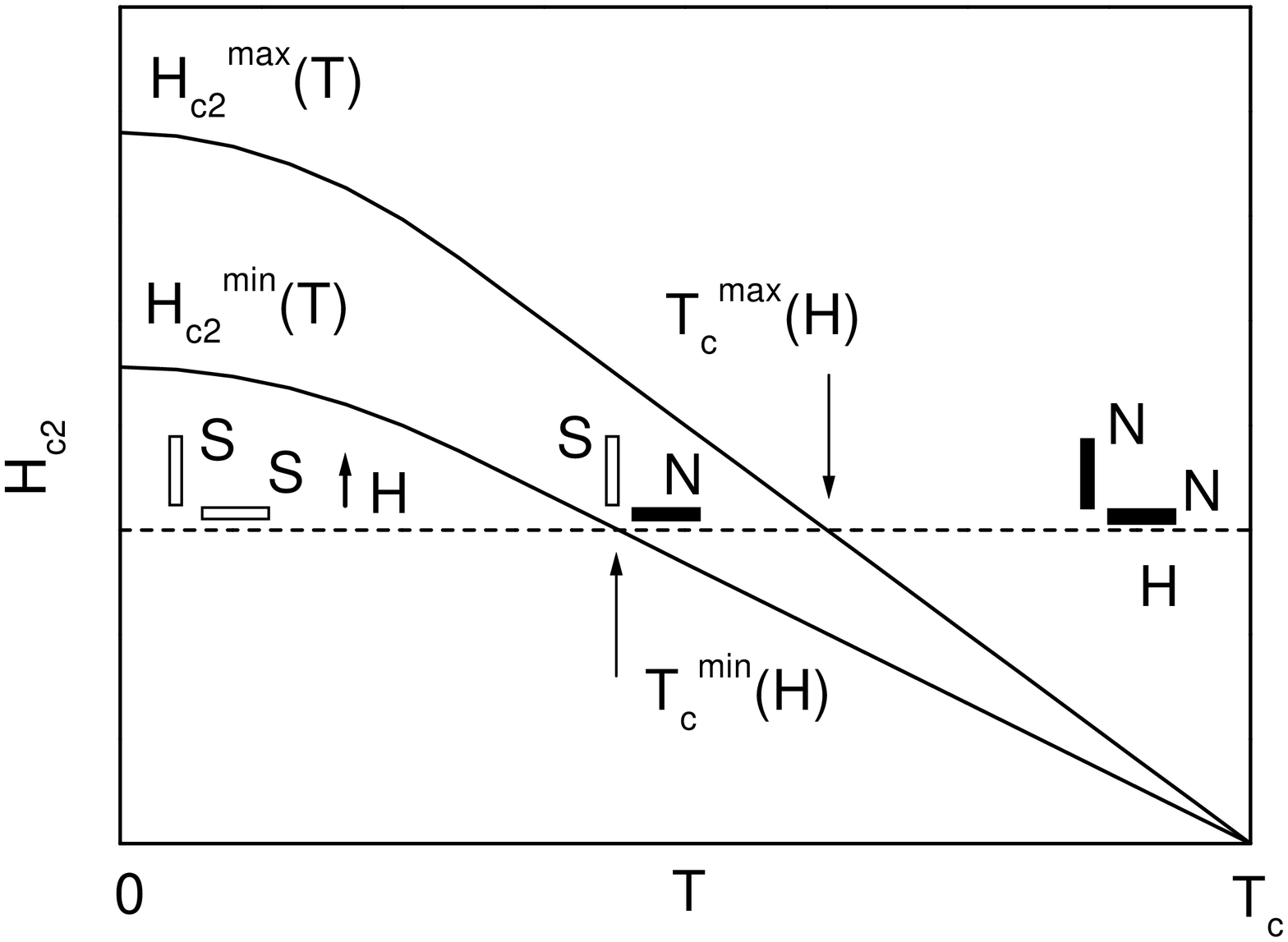}
}}
\caption{Sketch of the maximum  $H_{c2}^{max}(T)$ and the minimum
$H_{c2}^{min}(T)$ upper critical fields. For a given applied field
$H$, the relation
$H = H_{c2}^{min}(T_c^{min}) = H_{c2}^{max}(T_c^{max})$  defines
temperatures $T_c^{min},T_c^{max}$. The open (shaded) rectangles
represent superconducting (normal) grains for
   $T < T_c^{min}$, $T_c^{min} < T < T_c^{max}$, and $T > T_c^{max}$. }
\label{sketch}
\end{figure}
\pagebreak
\pagebreak
\pagebreak
%
\end{document}